# Brillouin imaging for studies of micromechanics in biology and biomedicine: from current state-of-the-art to future clinical translation

Christine Poon[1,2], Joshua Chou[2], Michael Cortie[1] and Irina Kabakova[1*]

[1] School of Mathematical and Physical Sciences, Faculty of Science, University of Technology Sydney, Ultimo 2007, Australia

[2] School of Biomedical Engineering, Faculty of Engineering and IT, University of Technology Sydney, Ultimo 2007, Australia

*corresponding author: irina.kabakova@uts.edu.au

**Abstract:** Brillouin imaging is increasingly recognized to be a powerful technique that enables non-invasive measurement of the mechanical properties of cells and tissues on a microscopic scale. This provides an unprecedented tool for investigating cell mechanobiology, cell-matrix interactions, tissue biomechanics in healthy and disease conditions and other fundamental biological questions. Recent advances in optical hardware have particularly accelerated the development of the technique, with increasingly finer spectral resolution and more powerful system capabilities. We envision that further developments will enable translation of Brillouin imaging to assess clinical specimens and samples for disease screening and monitoring. The purpose of this review is to summarize the state-of-the-art in Brillouin microscopy and imaging with a specific focus on biological tissue and cell measurements. Key system and operational requirements will be discussed to facilitate wider application of Brillouin imaging along with current challenges for translation of the technology for clinical and medical applications.

## 1. Introduction

Brillouin imaging (BI) is an emerging field that holds the potential to enable unparalleled mechanical mapping of cells, tissues and organs across three dimensions with submicron resolution [1-4]. It is rapidly gaining traction in biology and biomedical science communities due to its non-contact and label–free nature, offering clear advantages over existing methods of measuring mechanical properties on a nano to microscopic scale such as atomic force microscopy (or nanoindentation), optical tweezers and rheology (Fig.1). The first Brillouin images of cells were reported in 2015 [2] and inspired a growing number of studies in mechanobiology [5, 6], disease screening [7-9] and cancer research [10] where there is much unknown regarding cell biomechanics-function relationships. The benefits of *in vivo* BI in application to ophthalmology has already been demonstrated based on clinical studies with over 200 patients enrolled to date [11, 12]. *In vitro* studies have presented compelling evidence towards benefits of BI for detection of early changes in articular cartilage due to osteoarthritis [13] and potential use of BI as a diagnostics tool in atherosclerotic development of arterial plaques [14]. In principle, BI could be extended to the study of other diseases in which tissue mechanics are implicated. For example, both systemic fibrosis (an autoimmune disease in which excessive amounts of stiff extracellular matrix accumulate around organs [15]) and endometriosis (where endometrial tissues grow beyond the uterine cavity to form fibrotic tissues around organs [16, 17]) are poorly understood or underdiagnosed conditions where new approaches for early diagnosis or treatment are urgently needed.

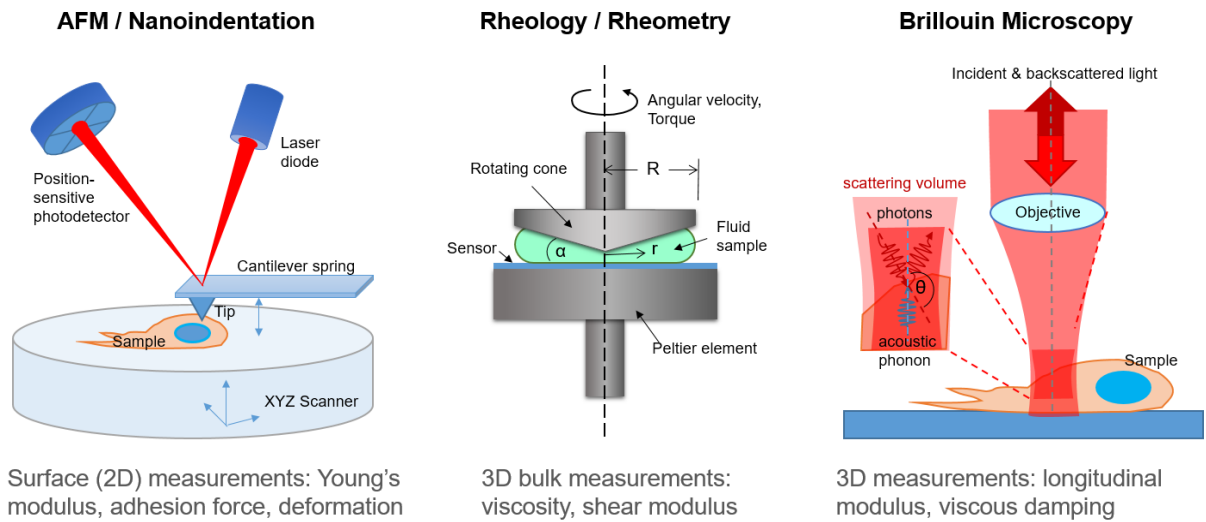

**FIGURE 1.** Comparison of techniques for measuring the mechanical properties of biomedical materials. AFM/nanoindentation and rheology require sample contact which can be destructive whereas Brillouin microscopy is based on light-phonon interactions and so is non-destructive by nature.

BI is based on the principles of Brillouin light scattering (BLS), an inelastic scattering of light by thermally-excited pressure waves. The phenomenon has been known since the 1920s due to independent discoveries by Leon Brillouin [18] and Leonid Mandelstam [19]. The original application of BLS was in the material sciences, enabled by the invention of the laser in the 1960s to become an established technique, known as Brillouin spectroscopy, for characterizing condensed matter in solid-state physics, crystallography and geology [20]. The last decade has seen a renaissance of Brillouin spectroscopy particularly in biological and medical imaging applications [21]. The two primary reasons behind for the recent revival of interest in BLS are: (i) the technology required to perform BLS measurements has drastically improved in the last 30 years with the invention and optimization of Sandercock-type interferometer [22], virtually imaged phase array (VIPA) spectrometer [1] and coherent nonlinear detection schemes [9, 23-26], and (ii) the growing field of mechanobiology borne from understanding that mechanical factors are important in the development and functioning of cells and disease pathogenesis [27, 28].

Despite significant improvement in the performance of BI systems and the plethora of applications where this technology can be transformative, BI is currently at a nascent stage that remains within the domain of academic research. Given the increasing interest in BI as a technique for gaining new insights in biology and recognition of its potential for medical imaging and diagnostics, here we provide a comprehensive summary of progress in the application of BI in biological studies for the benefit of interested researchers, and importantly, discuss key requirements and challenges necessary to overcome along the pathway towards its standardized and widespread clinical application.

## 2. Overview of Brillouin light scattering phenomenon and methods

### 2.1 General principles of Brillouin light scattering

Two processes occur when a beam of light passes through matter, namely, absorption and scattering. Atoms and molecules in matter absorb light energy, whereby some is lost as heat and the rest is re-emitted such that the light intensity decays exponentially with the propagation distance. Scattering phenomena depend on the size of a particle in relation to the wavelength of the incident beam and can be classified as either elastic (Rayleigh), where the direction of the beam changes but the frequency of the beam is conserved, and inelastic scattering (Brillouin and Raman), where the frequency and the direction of the beam is altered during propagation [29].

BLS is based on the interaction between photons and acoustic phonons, thermally excited vibrations within a material [18, 19]. As phonons are density perturbations that move at the speed of sound, such an interaction results in a Doppler shift of the scattered light frequency by precisely the phonon frequency $\Omega$. Thus, if the incoming photon has a frequency $\omega_i$, the scattered photon frequency is given by

$$\omega_s = \omega_i \pm \Omega. \tag{1}$$

The sign "±" suggests that the energy can either be transferred from phonon to photon ("+") or vice versa ("-").

The phonon frequency $\Omega$, or so-called Brillouin frequency shift (BFS), is the first measurable output of a BLS experiment. It is directly proportional to the acoustic velocity $V$ as

$$\Omega = \pm Vq = \pm \frac{4\pi n}{\lambda} V \sin\frac{\theta}{2}, \tag{2}$$

where $q = \frac{4\pi n}{\lambda}\sin\frac{\theta}{2}$ is the momentum exchanged in the scattering process, $n$ is the material refractive index, $\lambda$ is the wavelength of light in vacuum and $\theta$ is the scattering angle. The acoustic velocity, in turn, is the function of material density $\rho$ and the longitudinal modulus $M$

$$V = \sqrt{M/\rho}. \tag{3}$$

Eqs. (2) and (3) suggest that, in the assumption of constant $n$ and $\rho$, the data collected from BLS experiments can be used to assess $M$, also known as the constrained modulus. $M$ is one of the elastic moduli and is defined as the ratio of axial stress to axial strain in a uniaxial strain state. It should not be confused with the Young's modulus $E$, also called unconstrained modulus. In fact, in hydrated biomaterials $E$ and $M$ differ by several orders of magnitude and are not always correlated [30]. This has been previously discussed in a number of seminal works [3, 31].

The BLS experiment involves collection of spectral data with a schematic example of water and hydrogel spectra illustrated in Figure 2A and 2B. The spectra are centered on the frequency of the probing laser and consist of a number of peaks: a Rayleigh peak (R) at zero frequency and a Brillouin doublet, Stokes (S) and anti-Stokes (AS) peaks at frequencies given by Eq. (2). For most biological materials $\Omega$ is within the range of 1-10 GHz (for a visible laser probe) and it can be significantly higher in solid-state materials. The ratio of Rayleigh ($I_R$) to Brillouin ($I_B$) peak intensities is a constant for a given material, and is given by the Landau-Placzek ratio [32]

$$I_R/2I_B = (\beta_T - \beta_S)/\beta_S \tag{4}$$

with $\beta_T$ and $\beta_S$ being isothermal and adiabatic compressibilities, respectively.

The second measurable output of a BLS experiment is the full width at half maximum (FWHM) of the Stokes and anti-Stokes peaks, $\Gamma$. This parameter is inversely proportional to the acoustic lifetime $\tau \sim 1/\Gamma$, so the materials with larger acoustic damping show broader Brillouin peaks.

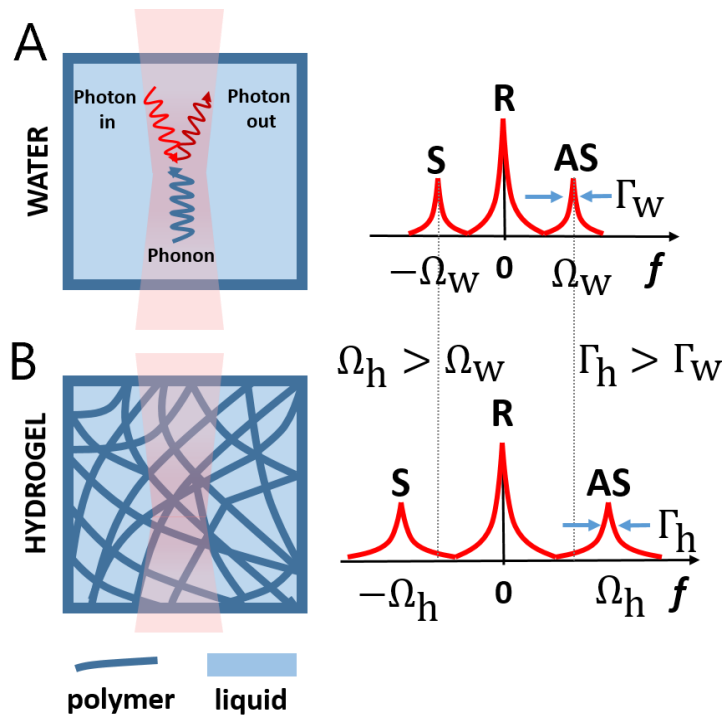

**FIGURE 2.** Schematic representation of the interaction between light and sound in media of varying complexity: (A) water and (B) hydrogel. The Stokes (S) and anti-Stokes (AS) peaks measured in water have smaller full width at half maximum ($\Gamma_W$) and are positioned closer to the central Rayleigh (R) peak. Broadening ($\Gamma_W < \Gamma_h$) and spectral shift of Stokes and anti-Stokes peaks in hydrogel ($\Omega_W < \Omega_h$) reflect the multicomponent and viscoelastic nature of hydrogel consisting of a polymer fiber network immersed in liquid.

The exact values of $\Omega$ and $\Gamma$ are determined not only by frequency and lifetime of phonons traveling inside the material, but also by the scattering geometry, properties of the imaging objective lens, and the sample's internal architecture. Overall, there are three spatial scales to consider: i) the phonon length, ii) the phonon propagation distance and iii) the imaging volume [33]. Depending on these three parameters and how they compare to one another, different effects can be observed. For example, if the internal structure of a multicomponent material has spatial scale much smaller than the phonon length, the frequency of acoustic phonon and its life time are represented by the average properties across the imaging volume. Consider a hydrogel material illustrated in Fig. 2B: the dimension of polymer fibers that construct solid network of hydrogel is on the order of 20-50 nm, whereas the phonon length is 200-300 nm and the imaging volume is typically 1-10 µm³. The Stokes and anti-Stokes peaks that will be measured in this scenario are broadened due to inhomogeneity and viscosity of the gel, and shifted toward the higher frequency (Fig. 2B), in comparison to the Brillouin scattering signal of pure liquid such as water (Fig. 2A). In general, interpretation of $\Omega$ and $\Gamma$ is not a trivial matter and largely depends on the type of material and the material model being adopted. We will discuss some of the models that find use in 'bioBrillouin' community in Section 5.

## 2.2 Terminology of methods based on Brillouin light scattering

As BI is becoming more established with more researchers of different backgrounds joining the field, there is a compelling need for unity regarding the use of terminology and clarity between different concepts, methods, technologies and descriptions of Brillouin applications that are often used interchangeably and/or mixed in the current literature. To address this need, we identify and define current techniques and technologies in the field and propose their classification under the umbrella term of Brillouin imaging for clarity. Our objective is to provide a guide to BI terminology which can serve as guidance for the growing Brillouin community.

*Brillouin imaging* is an umbrella term that encompasses all methods that employ BLS as a contrast mechanism to generate an image or 'map' of the sample of interest, whether by spectroscopy, microscopy or a combination of various techniques. We encourage adoption of this term whenever the technical details of a method are secondary in significance to the goal of a narrative and elaborate with the appropriate technique.

*Brillouin spectroscopy* refers to the single point detection and analysis of Brillouin light scattering collected from a material of interest. The scattering volume is not restricted to any particular size in this application and is mostly chosen based on consideration related to the material structure and the scattering geometry.

*Brillouin micro-spectroscopy* refers to the detection and analysis of a Brillouin light scattering signal from a micro-sized volume of a material. This typically requires implementation of high-numerical aperture lenses that can lead to spectral broadening and distortions of peaks in the scattered light spectrum [34].

*Brillouin microscopy* (BM) generally refers to the combination of Brillouin spectroscopy with confocal microscopy. Due to the construction of a confocal microscopy system that typically uses epi-illumination, only scattered light propagating in the backwards direction is collected and passed through the microscope aperture. The spectroscopic analysis is then performed for each spatial location within 3D sample by scanning the object mounted on a 3-axis microscopy stage [4].

*Shear Brillouin light scattering microscopy* is the method adopted for mapping both longitudinal and shear elastic moduli within the 3D sample. The instrument strongly resembles a standard confocal Brillouin microscope, but the probing light is sent at an angle to the optical axis of the objective lens and enters the lens at the periphery [35]. The scattered light collected by the same lens consequently has two components that come from interaction with the longitudinal and the shear pressure waves, resulting in respective retrieval of longitudinal and shear elasticity metrics.

*Line-scanning Brillouin microscopy* refers to sub-division of Brillouin microscopy in which the signal is collected instantaneously from an entire line inside the sample, in contrast to point-by-point collection [36].

*Stimulated Brillouin scattering (SBS) microscopy* is the term adopted for a Brillouin microscopy method based on the SBS process in which inelastic scattering of the probe light is enhanced by addition of pump light. SBS experiments are typically based on coherent heterodyne detection using radio-frequency modulation formats and have benefits of improved scattering efficiency and faster acquisition time [23, 24].

*Impulsive Brillouin scattering (ISBS) microscopy* is an alternative to SBS and employs two crossing pump beams to generate a transient density grating at the point of intersection within the sample, whereby scattering from this grating is measured with a continuous wave probing laser beam [25, 26]. This modulation is recorded in the time domain and the phonon frequency associated with the modulation is extracted through Fourier transformation. In spite of complexity of the experimental setup, the technique offers the clear advantage of time-resolved spectroscopy. The latter feature has been acknowledged by an alternative name of this method as *Brillouin Imaging via Time-Resolved Optical Measurements (BISTRO)* [9].

*Brillouin endoscopy* – an approach to Brillouin imaging based on integration of the Brillouin spectrometer with the fiber probe for flexible and in-situ collection of the scattered light [37].

## 3. Technology & system requirements for BLS-based biological imaging

### 3.1 Brillouin hardware

A Brillouin imaging setup consists of several fundamental components including: (i) a laser, (ii) an imaging system, and (iii) a spectrometer or spectral detection instrumentation (Fig. 3A). The challenge in building such a system comes down to inefficiency of the spontaneous BLS process (1 photon in a billion) and a narrow range of frequencies detected in BLS experiments (0.1-20 GHz). Typical optical spectrometers based on prisms and gratings fail to provide sufficient spectral resolution. Specialized instruments designed for spectroscopic measurements in the MHz range, e.g. tandem Fabry-Perot interferometers or lock-in detection schemes common in microwave and radio-frequency physics are better suited for Brillouin spectroscopy applications. Long phonon lifetimes (microseconds) require frequency-stabilized, high-coherence light sources to probe photon-phonon interactions. Special filters may be necessary to remove unwanted background composed of stray light, Rayleigh scattered light, and Raman and fluorescence signals [38, 39]. A scientific grade camera or a photo-multiplying tube should be used to capture the relatively low number of backscattered photons generated in BLS. In addition, electron-multiplying sensors (EMCCD cameras) are sometimes necessary to achieve sufficient measurement sensitivity, especially for imaging in turbid media. Below we give examples and describe each component of a Brillouin imaging system in detail.

### 3.1.1 Laser

The spectral range of Brillouin light scattering (typically between 1-20 GHz and sub-GHz line widths), determines the choice of laser suitable for intended experiments. The spectral width of the laser emission should be well below the Brillouin linewidth (~1MHz or less) to avoid spectral broadening of Brillouin peaks and maximize inelastic scattering. The frequency stability and noise figure of the laser is equally as important for reliable detection of Brillouin signals and sufficient signal-to-noise ratio. Frequency locking to an external reference is often required to reduce temporal drifts of the laser frequency due to temperature fluctuations [21, 40].

Apart from the considerations above, the ultimate choice of the laser source is also determined by sample transparency. Brillouin imaging and microscopy of biological materials are ideally performed within the 'tissue window' i.e. the region of minimal light absorption and scattering of biological cells and tissue, which occurs between the mid-visible to near-infrared regions [41]. The UV and blue/green spectral regions are not suitable since light can be absorbed by DNA, melanin, fat, bilirubin, or beta-carotene. In the infrared region, water absorption becomes dominant and thus, such lasers are also not a good choice for BI. On the other hand, BLS is a dipole-radiative process, thus the scattering efficiency is proportional to $\lambda^{-4}$, so the signal intensity is significantly weaker at long wavelengths. Another important advantage of using shorter wavelengths for BI is the ability to achieve higher spatial resolution (proportional to $\lambda$).

Brillouin studies so far have used frequency doubled solid-state lasers, most using 532 nm wavelength of Nd:YAG laser [2, 4, 35], some using 561 nm [5, 13, 30, 42] and two using 671 nm [37, 38]. For *in vivo* studies, where photodamage is a serious concern, BLS experiments have used near-infrared wavelength (780 nm) employing semiconductor lasers and additional spectral purification elements [8, 23, 43, 44]. A recent study by Nicolic *et al.* [45] showed that the use of 660 nm wavelength enables live cells to be irradiated with 82 times more energy than at 532 nm, thus shortening the acquisition time and allowing approximately 34 times higher signal-to-noise ratio. Clear advantages are that measurements can be performed over longer periods of time with lower risk of photodamage, over larger fields of view with shorter point-to-point steps, and with improved precision.

### 3.1.2 Spectrometer
The biggest challenge of Brillouin imaging has been in achieving sufficient spectral resolution and sensitivity necessary to resolve a relatively weak Brillouin signal only a few GHz away from the frequency of the laser. Several strategies and spectrometer designs have been suggested to address this challenge, all of which will be discussed in the following section.

### Tandem Fabry-Perot interferometer

Spectroscopic measurements with MHz resolution have traditionally involved scanning interferometers, e.g. Fabry-Perot interferometers (FPI) [22]. The main optical element of an FPI is a set of two parallel mirrors separated by a free space of certain length, or so-called Fabry-Perot etalon. Depending on the distance between the mirrors, light travels through or is reflected by the instrument. By scanning the distance between the mirrors, the spectral selectivity is achieved within the instrument's free spectral range, the spectral distance between neighboring maxima in the transmission spectrum. A standard FPI has one etalon in its construction and it can be set up to have required resolution in the range of a few tens of megahertz. However, its sensitivity is not always enough to detect Brillouin signals in semi-opaque and turbid samples. The coupling of two synchronized Fabry-Perot interferometers results not only in increased contrast, but also helps to broaden free-spectral range of the instrument [46].

A modern tandem Fabry-Perot interferometer (TFPI) can achieve contrast of up to 150 dB and hence is ideally suited for Brillouin imaging of turbid samples. Originally designed by J. R. Sandercock in 1970s, the TFPI was recently improved by the addition of optical isolators, which provide an extra 50 dB of optical contrast compared to the original design [33, 47]. FPIs are versatile instruments with several adjustable parameters. For instance, the mirror spacing can be changed from 0.1 mm to 30 mm to tune the instrument's free spectral range and spectral resolution. Thus the same instrument can be used for Brillouin measurements of solid samples like quartz ($\Omega$~30 GHz) and soft samples such as hydrogels ($\Omega$~6 GHz at the laser wavelength of $\lambda$~600 nm) with equal resolving power.

The TFPI is the only commercialized instrument in the Brillouin imaging field, which contributes to its popularity for Brillouin spectroscopy and the spread of technology around the world. The only downside of a TFPI system is the time required to acquire the signal (typically 1 s or more). The minimum acquisition time is limited by the mirror scanning speed and cannot be significantly improved. Although acceptable for single point measurement, long signal acquisition times are prohibitive for scanning across 3D volumes; a high number of sampling points (>10,000) can take many hours, which is not suitable for live cell imaging.

### Virtually imaged phase array spectrometer

Initially introduced in telecommunications [48], a virtually imaged phased array (VIPA) can be used to spatially spread the frequencies of the scattered beam. This constitutes the second, non-scanning, type of spectrometer for imaging based on spontaneous BLS [1]. Single-stage and multi-stage VIPA spectrometers are both popular options for Brillouin measurements that require faster acquisition speeds than possible with TFPIs [7, 49]. A simple design of VIPA optical element, consisting of a glass etalon with reflection coatings and a transparent input window [48], offers sufficiently large angular dispersion to separate Brillouin Stokes and anti-Stokes peaks from the Rayleigh signal. The resolution of VIPA spectrometer is limited by the fabrication tolerances of the etalon and realistically it is on the order of 0.5-1 GHz. Typically this is not enough to resolve FWHM of Brillouin peaks (50-500 MHz) and the measured signal is represented by the convolution of the material linewidth and the instrument response function. This is a significant downside of VIPA-based Brillouin spectrometer, as this only permits the measurement of the BFS. This spectrometer has no movable parts and no need for synchronization, making it much cheaper than TFPI and significantly easier to install and operate.

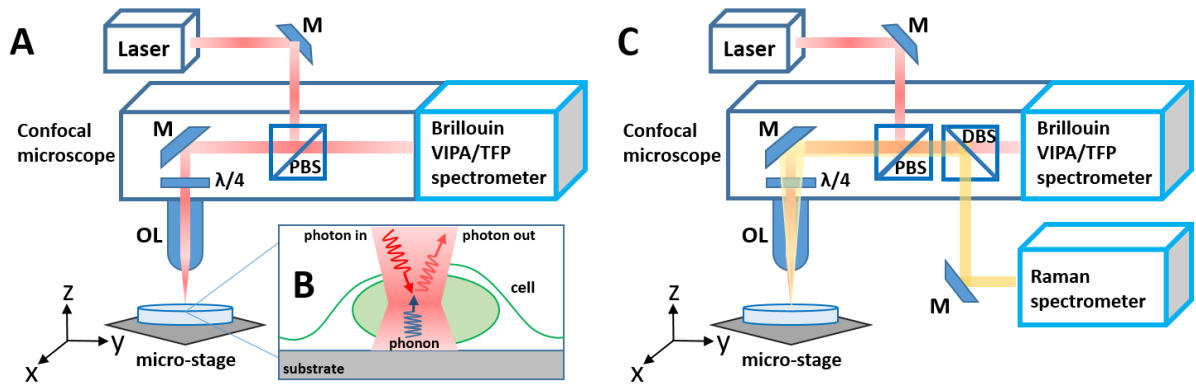

**FIGURE 3.** (A-C) Imaging systems based on spontaneous scattering. (A) Spontaneous confocal Brillouin microscope can utilize VIPA or TFP spectrometers. Notations used: M – mirror, OL – objective lens, PBS – polarization beam splitter, λ/4 – quarter-wave plate. (B) Photon-phonon interaction in a biological material such as cell. (C) Brillouin and Raman microscopy can be combined in one measurement system by splitting the scattered beams with a dichroic beam splitter (DBS) ensuring instantaneous mechanical and chemical read out from the same location within the sample.

A typical acquisition time of VIPA spectrometers is on the order of 100 milliseconds, depending on the sample transparency and laser power, which is at least an order of magnitude faster than TFPIs. The contrast available with single-stage VIPA spectrometer is, however, only 30 dB, making a single stage-system unsuitable for measurements of turbid media. Addition of the second VIPA etalon, aligned at 90-degrees to the first one, can boost the contrast of the spectrometer to 50-60 dB but with a reduction in signal strength. Overall, VIPA spectrometers are a good choice for transparent samples and quick measurements which aim to detect relative changes in the Brillouin frequency shift without a need for details of the spectra. The moderate cost of this system also suggests that it can be suitable for routine monitoring and point-of-care diagnostic environments. It is hence not surprising that the only clinical trials to date — diagnostics of ocular health — are based on Brillouin imaging using a VIPA-type spectrometer [8].

### Coherent nonlinear techniques

Nonlinear techniques such as stimulated Brillouin scattering (SBS) imaging [23, 24] and impulsive Brillouin imaging [25, 26] form an alternative group of Brillouin imaging methods that exploit formation of an acoustic wave by electrostriction or thermal excitation. The phenomenon of electrostriction is associated with the change in material density as the result of application of a strong electromagnetic field. Light absorption and thermal expansion could be the second route for excitation of phonons in the material. Regardless of mechanism, phonons generated as the result of exposure to high-intensity periodic optical fields can be probed by a weaker beam of light to obtain the phonon propagation speed and ultimately some of the material's mechanical properties. The efficiency of phonon generation in this case is proportional to the pump light intensity, and hence the probe scattering could be orders of magnitude stronger than in the scenario of spontaneous Brillouin scattering. High scattered signal magnitude translates into better signal-to-noise ratio of stimulated versus the spontaneous techniques, and consequently faster acquisition times.

Stimulated Brillouin imaging uses a strong pump and a weak probe beams, slightly detuned in frequency. By scanning the probe frequency, phonon generation reaches the maximum efficiency whenever the frequency detuning between pump and probe coincides with the acoustic resonance (corresponding to the frequency of one of acoustic modes supported by the sample's geometry and composition). The detection of the scattered signal is achieved via radio-frequency modulation formats and lock-in detection schemes [23]. Such schemes are better suited for resolving GHz frequency shifts with resolution only limited by the laser bandwidth (that can be below 1 MHz).

| BI Technique | BLS Type | Acquisition time (s) | Spectral Resolution (MHz) | Contrast (dB) | Sample Suitability | Power* (mW) | Refer. |
|---|---|---|---|---|---|---|---|
| VIPA | Spont. | 0.1-10 | 500 | 10-30 (1 stage) 30-50 (2 stage) | liquids hydrogels cells tissues organs | 1-30 | [2, 4, 5, 13, 40, 49, 50] |
| TFPI | Spont. | 1-100 | 10-100 | 100-150 | liquids cells tissues | 10-30 | [3, 10, 33, 40, 47, 51, 52] |
| SBS | Stimul. | 0.05-10 | 4 | 20-50 | water intralipid | 270 mW (pump) | [23, 24] |
| ISBS | Stimul. | 0.001-0.05 | 1 | 10 | methanol NaCl solution hydrogels | 20 mW (probe) 4-40 µJ (pump) | [9, 25, 26] |

*on sample

**TABLE 1**. Key parameters of Brillouin imaging techniques (SI unit abbreviations).

In the impulsive Brillouin imaging technique, the acoustic wave is formed by crossing high-power pump beams at the same frequency. The density grating built by electrostriction and thermal excitation then propagates through the samples and is probed by a weak probe beam (Fig. 4A). The scattered signal is detected via a fast photo-diode and the frequency is extracted by fast Fourier transform [25, 26]. This scheme enables assessment of the time-dependent dynamic processes, e.g. polymerization processes or environmental changes on millisecond time scales, due to the fast processing times and direct access to temporal characterization of the travelling acoustic waves (see schematic of the measurement system in Fig. 4B).

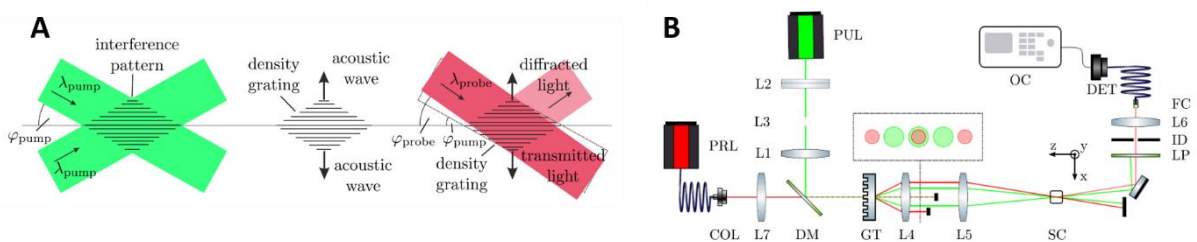

**FIGURE 4**. (A) Principle of Impulsive Brillouin Imaging microscopy with pump (green) and probe (red) beams. High-intensity pump pulses cause formation of a moving density grating through the thermal and electrostriction processes; a weak probe is scattered by the density grating resulting in a Brillouin frequency shifted signal. (B) Schematic of the setup: PUL excitation laser (pulsed 12 ps, 532 nm); L2 L3 cylindrical lenses; L1 L4 L5 L6 L7 achromatic lenses; PRL probe laser (CW, 895 nm); COL collimator; DM dichroic mirror (long-pass); GT grating; SC sample container; LP long-pass; ID iris diaphragm; FC fiber connector with long-pass; DET detector; OC oscilloscope. Materials reproduced with permission from the source [26].

*3.1.3 Imaging system*

Confocal microscope

Koski and Yarger were the first to combine Brillouin spectroscopy with confocal microscopy to achieve Brillouin imaging of heterogeneous samples [1]. The spatial resolution of this experiment (20 µm) was far from modern standards in confocal imaging, but later experiments pushed the spatial resolution to the diffraction limit and achieved subcellular imaging with the beam focused to dimensions below 500 nm [2, 5]. The epi-illumination principle common in confocal imaging, i.e. the objective used to illuminate the sample, serves for light delivery and collection (Fig. 3A and 3B). The measurement is set-up in a reflection mode and is suitable for 3D bulky samples and in situ imaging. The 180-degree scattering geometry, however, restricts the phase-matching condition between optical and acoustic waves, selecting only one type of interaction with longitudinal acoustic waves [4]. Thus, the information available in confocal imaging experiments is limited to the measurement of the longitudinal acoustic velocity and the life time of longitudinal phonons. Full characterization of mechanical properties necessitates other types of mechanical perturbations e.g. shear and Young's moduli.

### Platelet scattering configuration

Platelet scattering geometry allows measurement of anisotropy of the acoustic velocity by changing the angle of incidence and is suitable for characterizing mechanical properties in tissues with complex structure and directions of symmetry. A thin section of the specimen (100-200 µm) is positioned on a reflecting surface (a mirror). An incident light beam probes the specimen at a non-zero angle to the normal of the specimen's surface, while the mirror holding the sample is rotated 360 degrees around the normal. This scattering configuration allows access to the velocity of two types of acoustic waves, parallel to the mirror surface and perpendicular to it. It has been widely used to study anisotropy of elastic properties in collagen and elastin fibers [51], cartilage [53] and bone specimens [54].

### Fiber-integrated probes

The non-contact and label-free nature of Brillouin imaging makes this technology an ideal solution for *in vivo* and *in situ* imaging for a number of applications in biomedical or industrial monitoring. The footprint of a Brillouin imaging system and its reliance on bulk-optical alignment is, however, the major drawback in the translation of this technology into real-world environments. Despite the dramatic improvement in the performance of Brillouin microscopes since the first publication by Koski and Yarger in 2005 [1], much more effort is needed to integrate the instrument into a scalable, robust and easy to transport device. Some progress in this objective was made by Kabakova *et al.* with a study of the performance of Brillouin fiber probes [37]. Single and dual fiber probe designs were evaluated in terms of their collection efficiency and imaging performance, with dual fiber design offering a straightforward solution to the removal of unwanted fiber-generated Brillouin scattering background [37]. Further work is, however, needed to improve the efficiency of the scattered light collection in such fiber probes and to create monolithic fiber devices capable of 360-degree imaging within the tissue [55].

### 3.2 Multimodal and correlative systems

Application of BLS as a contrast mechanism to confocal imaging of biological samples by Scarcelli *et al.* technically represents the first combinative Brillouin microscopy system. Currently, systems that combine Brillouin microscopy with other established techniques are rapidly emerging, capable of providing specific quantitative analyses of the chemical and physical properties of samples with simultaneous viscoelastic mapping inferred from Brillouin imaging. These systems represent an important advancement in correlative microstructural characterization. By allowing direct correlation of distinct metrics, a more holistic characterization and elucidation of biological phenomena can be obtained.

### 3.2.1 Brillouin-Raman microscopy

Raman spectroscopy is an established technique for probing the molecular structure, composition and vibrational modes of compounds. Based on inelastic scattering, it can provide a unique 'fingerprint' that can be used for chemical identification. Recent advancements have enabled the integration of what were previously standalone systems to provide simultaneous Brillouin and Raman light scattering detection capable of correlating viscoelastic characteristics with chemical specificity (Fig. 3C). As cellular processes are governed by a plethora of complex chemical interactions, Brillouin-Raman systems hold significant potential for studying biological processes such as tissue remodelling in disease progression and wound healing, or in reaction to changes in physico-chemical environments in events or therapies. The capacity to correlate chemical specificity to viscoelastic modulus is also important for the characterization of hydrogel-based tissue engineering substrates where stiffness is determined by crosslinking and hydration.

Higher contrast and subcellular resolution imaging with chemical and mechanical specificity was first reported by Palombo *et al.* to map the Brillouin signatures of *ex vivo* Barrett's oesophagus tissue site-matched with chemical signatures by Raman and FTIR [56]. All three measurements were performed separately and thus required a significant effort in image processing and data analysis due to ambiguity in physical location identification and matching between the three sets of experiments. High-performance simultaneous Brillouin and Raman measurements, collected from the same imaging volume within the sample, have been achieved a few years later by the same group of authors [33, 47, 57], representing dramatic improvement in imaging and data post-processing times, as well as significantly simplifying the results analysis and interpretation.

### 3.2.2 Brillouin-fluorescence microscopy

Elsayad *et al.* engineered a microscope system that integrated fluorescence emission detection with detection of BLS and called the method fluorescence emission–Brillouin scattering imaging (FBi) [58]. The authors demonstrated that this approach can be used to investigate regulatory events that alter cellular and extracellular mechanical properties of living cells within tissues. This work also revealed that the cytoplasm near the plant cell membrane and the extracellular matrix are regions of locally increased stiffness and the stiffness is different along and perpendicular to the cell growth axis [58].

Another study integrated Brillouin micro-spectroscopy with microfluidics to achieve phenotyping of the cell nucleus at a throughput of 200 cells per hour [7]. Fluorescence imaging in this study was used to verify Brillouin signatures and separate signals associated with nuclei and cytoplasm.

Finally, the combination of Brillouin and fluorescence imaging was used to study the stiffness evolution and growth dynamics in active, live *P. aeruginosa* biofilms [59]. The spatial distribution of the fluorescence signal in this study helped to correlate the value of the Brillouin frequency shift with the mass distribution of live bacteria within the colony.

### 3.2.3 Brillouin imaging and optical coherence tomography

Tissue biomechanics during cranial neural tube closure in mouse embryo development was recently measured *in situ* using combined Brillouin imaging and optical coherence tomography (OCT) [60]. OCT, an optical imaging technique that does not require physical contact or labels, was perfectly suited to guide the experiment and provide structural information on the stages of the cranial tube closure. The mechanical contrast supplied by Brillouin imaging at submicron resolution showed gradients in the longitudinal modulus across the cranial tube and tube stiffening at later stages of embryo development [60].

### 3.3 Sample preparation

A key limitation for optical microscopy is that biological samples beyond a thickness of several cells tend to be opaque due to the cellular and inhomogeneous structure of tissues. As Brillouin spectroscopy is fundamentally based on light scattering and acoustic wave interactions within a material, sample opacity limits penetration of the probing laser beam and produces internal scattering, which confounds Brillouin signals and measurement. A strategy to improve sample transparency for optical microscopy is known as optical clearing. This uses a combination of techniques to homogenize the refractive index of the overall sample. These include: delipidation, the removal of the lipids that contribute to opacity using solvents, urea or detergents; dehydration, hyperhydration or 'refractive index matching', whereby samples are soaked in high refractive index solutions that match the average refractive index of most biological tissues (between 1.44-1.52) [61]; or hydrogel embedding, where samples are incubated in a hydrogel and proteins are crosslinked prior to delipidation to preserve their structure [62].

Removal of structural components and equilibration of tissue with solvents and polymeric liquids can be expected to cause significant changes to mechanical properties of a sample. Riobioo *et al.* investigated the effect of optical clearing of rat brain and heart tissues on Brillouin spectra and, surprisingly, observed only subtle differences between the BLS spectra of cleared and uncleared samples despite clear changes in the overall structure and morphology of the organs after clearing [63]. As mechanical properties are intrinsically linked to material composition and structure, further studies are recommended to confirm the compatibility of sample clearing with Brillouin microscopy and clarify the interpretation of results. It is also recommended that any sample preparation procedure that may affect the structural mechanics of samples in any way be thoroughly investigated and reported, including how samples are extracted, processed or mounted for Brillouin analysis. In an exemplar, Edginton *et al.* provide a detailed protocol on how extracellular matrix (ECM) protein fibers were extracted and prepared for Brillouin measurement [64]. In the current status of the field, such methodology studies are necessary for meaningful comparison, evaluation of techniques and interpretation of results.

## 4. Progress in imaging cells, tissues and biological systems

Biological samples are inherently heterogeneous, structurally complex on a microscopic level and affected by temporal processes. It is increasingly recognized that mechanical properties of cells and tissues are integral and indicative of their function. However, as previously introduced, there has been a lack of appropriate means to measure mechanical information on a microscopic level across three dimensions. Brillouin spectroscopy combined with confocal microscopy has propelled the application of BLS for probing the viscoelastic properties of biological systems across a range of complexities, from single cell and subcellular components to whole tissues (Fig. 5 schematic), with the aim to become an established technique for better understanding structure-function relationships, fundamental cell biology, disease pathology and the growing field of mechanobiology. In [6], Prevedel *et al.* discussed considerations for the application of BM for elucidating cellular responses to the mechanical properties of their immediate environments as well as for meaningful interpretation of Brillouin scattering signals. In this section, we summarize and discuss key developments in the application of BI for measuring the viscoelastic properties of biological samples and systems. The following section is presented in terms of both hierarchical complexity of samples studied to date, from ground matrix and supportive ECM to cells, tissues and organs, as well as historical progress based on technological capability.

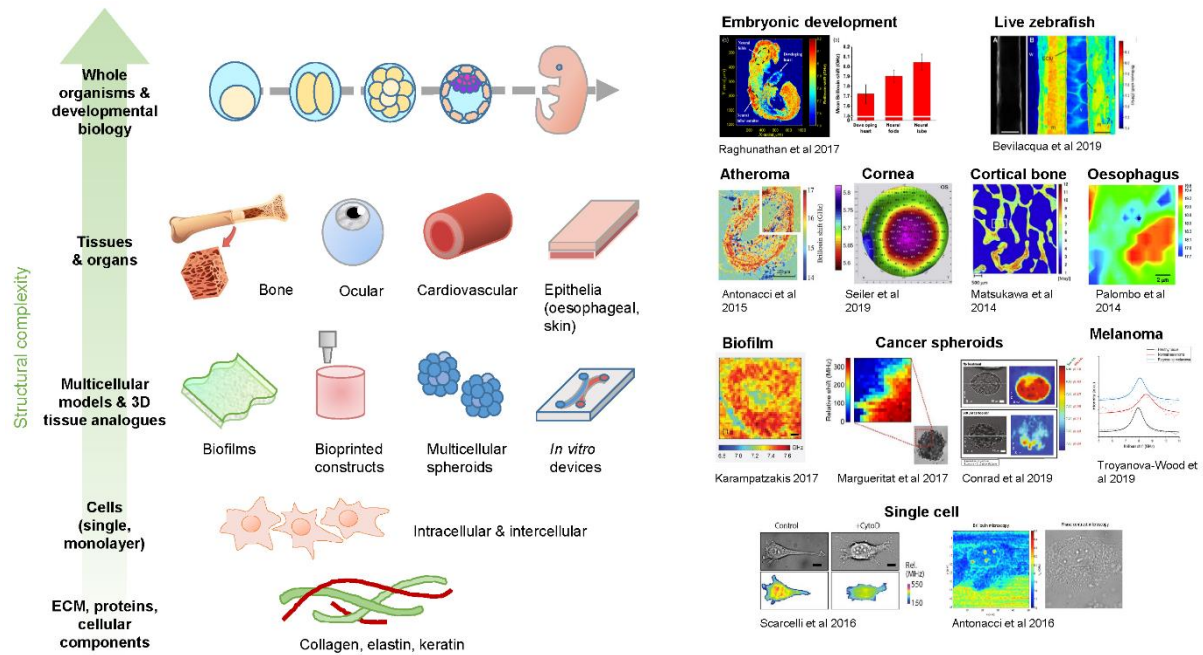

**Figure 5**. Left: schematic overview of biological systems that can benefit from BLS analysis organized according to structural complexity. Right: representative Brillouin-based imaging results collated from studies utilizing BLS for analysis of biological samples to date.

## 4.1 Cellular and tissue components

Biological tissues are composed of organized groups of structurally and functionally similar cells held together by ECM. While ECM composition varies between tissues, the three major constituents are: (i) insoluble collagen fibers which provide structural strength, (ii) viscous proteoglycan proteins which cushion cells, and (iii) soluble adhesion proteins which bind proteoglycans and collagen fibers to cell receptors. The mechanical properties of the ECM of biological tissues are dependent on their constituent biopolymers and important for normal tissue function, where disturbances in these properties are key biomarkers of disease [28, 65]. Furthermore, the composition and organization of ECM components gives rise to unique structural-functional dependent mechanical properties of each tissue and therefore represents a key biological structure of interest in the study of tissue biomechanics.

Collagen is the most abundant structural protein of the ECM and collagen fibrils are the smallest load-bearing structural elements of tissues. Studying the mechanical properties of collagen in isolation can enable better approximation and modelling of the properties of the source tissue(s) of interest particularly if the composition of collagen is known or can be mapped. Indeed, the first application of BLS for biological analysis was described by Harley *et al.* in 1977 in an attempt to correlate the molecular structure of rat tail collagen fibers to elastic moduli derived from Brillouin spectral measurements [66]. In this work, the elastic moduli of tropocollagen molecules in dry collagen were calculated using other known metrics such as density, peptide molecular weight and an estimate of hydrogen bond force constants. Harley *et al.* then examined the effect of hydration on measurements of dry collagen and reported hysteresis and lower measured modulus with hydration of the collagen fibrils as well as an inverse relationship between the degree of hydration and sound velocity. The study also showed that the high-frequency micro-Young's modulus of collagen fibers is an order of magnitude greater than that obtained from the Hooke's law region of a stress-strain curve, suggesting frequency-dependent molecular-scale viscoelastic properties. Cusack *et al.* extended these findings to include measurements of the anisotropy of collagen fibrils by measuring the velocity of propagation of longitudinal and transverse polarized elastic waves at different angles to the fiber axis. These were combined to determine the elastic constant of collagen [67].

Similarly, Randall *et al.* sought to determine the effect of fiber orientation with BFS in wetted rat tail collagen and horse hair keratin [68]. Differences in peaks with scattering vectors parallel, perpendicular and at a 45° angle with the length of the fiber were demonstrated, with implications in fiber structure and probing direction on Brillouin measurements. Lees *et al.* similarly measured BLS across wetted collagen and various dried and/or mineralized tissues including antler (bone), tibia and tendon and compared sound velocity resolved into axial and radial components for each material [69]. Optical refractive index of mineralized samples was found via Brillouin scattering. While mineralization is not standard practice for tissue sample preparation for biological characterization, the technique enabled sample compatibility with the optical systems of that time.

More recently, Palombo *et al.* characterized dehydrated collagen and elastin, major constituent proteins of ECM, using micro-focused BLS and reported Young's moduli values of approximately 10 and 6 GPa for collagen and elastin respectively [70] which are in line with previous findings. Advances in combinative systems have enabled more detailed correlative analysis. In [57], Mercatelli *et al.* describe application of BLS, Raman spectroscopy and second-harmonic generated maps in combination to examine sutural lamellae, collagenous structures within the cornea, in order to determine structure-function relationships in healthy and diseased states.

In addition to ECM mechanics, the presence and properties of pathological proteins are implicated in certain diseases and ageing processes. Antonacci *et al.* studied the biomechanical properties of stress granules containing FUS proteins, implicated in amyotrophic lateral sclerosis (ALS), within whole cells using a unique background-deflection Brillouin (BDB) system [71] and found an increase in stiffness and viscosity of stress granules containing ALS mutated FUS proteins.

It is expected that collateral advances in hardware capability, experimental methodology and signal analysis will enable finer discretization and resolution of proteins and ECM components within living cells and tissues. In [72], Bevilacqua *et al.* achieved sub-micron resolution in application of BLS to map changes in the mechanical properties of ECM in live zebrafish during development on sub-micron resolution, highlighting the importance of selecting optimal imaging parameters and spectral analysis for successful application of Brillouin microscopy (BM) as well as demonstrating the potential of BLS as a valuable tool for elucidating the role of ECM biomechanics in developmental and mechanobiology.

### 4.2 Biological systems
Here we focus on Brillouin analysis of *in vitro* multicellular models, which encompasses cell cultures in an increasing level of complexity from single cell to multicellular models, as well as relevant tissue-analogue systems.

#### 4.2.1 Single cells
The cell is the fundamental unit of all living organisms. The ability of modern cell culture techniques to sustain cells *ex vivo* and grow them in a controlled manner has underpinned a large part of modern medical research. Brillouin analysis of single cells is currently still at an early stage however. At the time of this review, reports in the literature have primarily been for proof-of-concept and spatial resolution testing during system development. The primary aim has been to demonstrate successful resolution of subcellular structures with ever improved clarity, which is essential for enabling detailed biomechanical studies of cells on submicron levels. It is expected that much can be yet discovered from a single cell to organ level using the continuously improving variations of BM.

In one of the earliest applications of BM to single cell analysis, Scarcelli *et al.* sought to determine the sensitivity of BM to osmotic factors, which affect the water content and organization of cytoplasmic structures within cells and therefore measured stiffness [73]. In [2], subjecting NIH 3T3 cells to hyperosmotic shock by increasing sucrose concentration to culture media was shown to induce a significant increase in stiffness measured across individual cells immediately after sucrose treatment, with a linear increase in stiffness with increasing sucrose concentration; these findings implicate consideration of the ionic and osmotic conditions of culture in experimental design and interpretation of BM measurements.

The nucleus is the largest and most rigid structure within eukaryotic cells, and therefore its physical properties contribute critically to the overall properties and mechanical behavior of cells. The structure and mechanical properties of the cell nucleus are known to regulate gene expression and key cellular processes such as proliferation, migration and differentiation [74]. Nuclear mechanics vary throughout the cell cycle and are important indicators of healthy and diseased states [75], but are very challenging to measure directly using physical means such as AFM. The non-contact nature of BM has positioned it as an advantageous tool for detailed study of nuclear mechanics. With sufficient resolution, BLS can be an effective contrast mechanism for differentiating nucleus from cytoplasm. In [5], Antonacci *et al.* successfully resolved nucleoli from the nuclear membrane from cytoplasm of porcine aortic cells. The work also measured a 3.6% decrease in cytoplasmic stiffness (to 2.51 ± 0.03 GPa) in cells exposed to latrunculin-A, a drug that lowers cell stiffness by preventing cytoskeletal assembly, verifying the suitability of BM as a non-contact and non-destructive technique for measuring nuclear and intracellular mechanics.

Combined or correlative systems extend simple Brillouin analysis by providing an additional metric for mapping properties across cells. Mattana *et al.* also differentiated nucleus from cytoplasm, in this case using a Brillouin-Raman system which correlated viscoelastic properties with biochemical composition. A significant reduction in cell stiffness was measured following transfection and oncogene expression [33], supporting the technique as an early biomarker for cancer detection. As mentioned earlier, Elsayad *et al.* have engineered an integrated fluorescence emission-Brillouin imaging configuration along with its namesake technique (FBi), which is shown to enable simultaneous visualization and mechanical mapping of structures within cells at submicron resolution [58].

The ability to measure changes in the stiffness across cells in a contact-less, label-free manner opens up a plethora of Brillouin analysis-based applications for study of cellular biomechanics in disease and therapeutic development. Zhang *et al.* demonstrate the extension of Brillouin microscopy to a flow cytometry technique capable of classifying cell populations based on nuclear mechanical signatures when flowed through a microfluidic device [76]. In [7], treatment of fibroblast cells with trichostatin-A, a chromatin decondensation drug, were shown to cause a significant reduction in nuclear stiffness, allowing successful differentiation from populations of untreated cells. It is expected that as BM becomes established as a technique for studying the biomechanics of cells and tissues, more applications will emerge, e.g. complementing single-cell omics.

*4.2.2 Spheroids and organoids*

Beyond a single cell, cell spheroids represent a more realistic *in vitro* tissue model compared to 2D culture and are currently the gold standard for therapeutic screening. Spheroids are clusters of cells induced to grow in a spherical formation by physical manipulation techniques, e.g. round bottom microwell plates, hanging droplet culture and encapsulation [77-80]. These cell aggregates represent micro-tissues and their 3D nature presents challenges for mechanical characterization and optical imaging throughout the entirety of a spheroid while simultaneously preserving its functionally relevant structure, presenting yet another opportunity for BM. Margueritat *et al.* explored application of BLS for mapping across cancer spheroids and demonstrated a clear delineation between the looser peripheral layer of cells and stiffer core of a spheroid [81]. Similarly, Conrad *et al.* measured the Brillouin frequency shift range of an ovarian cancer model cultured within a 3D gel matrix (Matrigel), as spheroids formed in a standard low-adhesion round-bottom plate as well as investigated the effects of hyperosmotic and carboplatin treatment [82]. Increasing osmolality was shown to increase Brillouin frequency shift measured from the cancer nodule cultures, corroborating previous findings [2], while carboplatin, a chemotherapeutic agent, was shown to decrease Brillouin frequency shift in treated groups due to disruption of the structural integrity of the nodules. It is expected that scanning capability coupled with improved signal deconvolution and 3D volume rendering will enable full realization of BM for detailed study of spheroids and organoid *in vitro* models as well as their responses to therapeutic agents in drug screening.

### 4.2.3 Tissue analogue models

The field of tissue engineering aims to recapitulate functional biofidelic tissues and organs using a combination of biomaterials, cells and relevant factors. Also known as tissue constructs, tissue analogues and tissue phantoms, engineered tissues currently serve as a more ethical and readily producible platform for drug screening, disease modelling, or for patient-specific organ replacement. While many methods to fabricate 3D engineered tissues exist, bioprinting is the most recent development in enabling the generation of complex structures. While a variety of biomaterials are compatible with bioprinting technologies, hydrogels have been the most commonly investigated due to their biocompatibility and proximity of their physical characteristics with ECM. These hydrogels, also known as 'bio-inks', are composed of natural, synthetic or composite biomaterials that may be chemically or crosslinked via conjugation with a photo-initiator to form controlled structures. Being soft and viscoelastic, hydrogel mechanics are difficult to characterize conventionally through rheology which only allows assessment of bulk volumes. Given that cells are sensitive to local environmental cues, there exists an immediate opportunity to utilize BM for studying and characterizing properties of bioprinted hydrogel constructs on a microstructural level that is directly relevant to the cells. This has the potential to provide more conclusive insights into cellular responses and construct remodelling processes. In [83], Correa et al. demonstrated an efficient Brillouin data collection and image analysis workflow on collagen gelatin hydrogels with different stiffness created by varying crosslinker concentration and formalin concentration. This is valuable for analysis of biologically-relevant tissue analogues with physiological hydration levels and subjected to routine fixation methods.

### 4.2.4 Biofilms

Biofilms are an organized collective of single-cell microorganisms of the same or different types that cohesively coexist within a shared self-excreted ECM sheet or membrane, wherein cells are attached to each other and frequently on surfaces. Microorganisms that typically form biofilms include bacteria, fungi and protists; a common example of a biofilm in the body is dental plaque. Biofilms also form on medical implants, can pose significant health risk, and are difficult to remove. Understanding the mechanical properties of biofilms will enable development of treatments that disrupt the ECM and stability of the biofilm, preventing its formation or facilitating removal from implant surfaces. Karampatzakis *et al.* used BM to measure the internal stiffness of live *Pseudomonas aeruginosa* biofilms under continuous flow and found that stiffness tended to increase towards the center of smaller colonies (films), indicating higher structural complexity akin to a spheroid in 2D [84]. Extension of BM with endoscopic techniques may enable real-time *in situ* study of biofilm formation on medical implants.

### 4.3 Tissues and organs

The initial success of BLS-based single cell studies sparked rapid progression to Brillouin imaging of tissues and more complex biological systems. Tissues are organized groups of similar cells that together, perform specific functions. There are four main types of tissues within the body: epithelial, connective, muscular and neural, whereby the composition, organization and orientation of cells defines each tissue and their unique functions. Organs are comprised of two or more tissues that operate in an anatomically distinct unit. Here we summarize Brillouin studies of biological samples within the structural complexity range of tissues to whole organisms.

#### *4.3.1 Cornea and eye*

The cornea is the clear frontal component of the eye which acts as a protective cover for the iris, pupil, lens and anterior chamber of the eye. The cornea is composed of water, collagens and glycan proteins. Likewise, the lens is also composed of collagens and glycosaminoglycans, with a uniquely crystalline structure that focuses light onto the retina as the basis of the mechanism of sight. Being transparent and therefore possessing high optical clarity as requisite for optical imaging, the cornea and lens are among the first tissues measured with Brillouin imaging systems. Randall *et al.* were among the first to report Brillouin spectral measurements of the cornea and lens of the human eye along with those of a variety of vertebrate species [85, 86].

Scarcelli and Yun described the first *in situ* Brillouin measurement of the lens within a mouse eye [4] and stiffness map of *ex vivo* cornea [49] to set the precedent for subsequent studies of ocular tissue mechanics using Brillouin microscopy. Distinct differences in biomechanical properties between normal corneas and keratoconus were measured via a line-scanning Brillouin setup [8]. Given the high water content of the cornea, Shao et al determined the sensitivity of BM in detecting changes in hydration level within corneal tissues to facilitate analysis and interpretation of results [44]. In [87], Scarcelli et al applied BI to assess the efficacy of corneal stiffening via collagen crosslinking, a therapeutic means to halt the progression of keratoconus and corneal ectasia, on porcine cornea. Together, these results support the establishment of a number of clinical trials to evaluate and validate BI as a tool for diagnosing ocular anomalies [11]. In [12], Shao et al describe the first clinical trial and protocol using BI to map across corneas *in vivo*, where increasing biomechanical inhomogeneity was found to occur within the cornea with progression of keratoconus, as well as asymmetry in mechanical properties between the left and right cornea at the onset of the condition. This study was then extended, where Seiler et al measured a significantly lower Brillouin frequency shift in the thinnest regions of keratoconus corneas compared to normal corneas ($5.7072 \pm 0.0214$ vs $5.7236 \pm 0.0146$ GHz, $P < .001$) [88].

While promising, clinical BLS measurements of corneal or ocular tissue necessitates consideration and resolution of two important issues: (i) the intrinsic risk of laser irradiation damage to the eye and (ii) motion-induced vibrational noise. More advanced optical technologies able to provide improved signal to noise ratio will enable a safe operational laser power compatible with live measurements of ocular tissues. In addition, future improvements to signal detectors and the development of more sophisticated processing algorithms will be able compensate and filter out natural movements and vibrations. These considerations will be further discussed in Section 6.

### 4.3.2 Bone and cartilage

Bone and cartilage are specialized forms of connective tissue that provide structural support for muscles and organs within the human body. Bone is hard, highly vascularized and calcified connective tissue that comprises the skeleton whereas cartilage is soft, flexible, avascular, tissue that serves to absorb shock in the joints between bones, or forms rigid, low-load bearing structures such as within the nose, ear and larynx. The structure and mechanical properties of bone and cartilage are integral to their function, and disruption of these properties are implicated in injury and degenerative conditions such as arthritis and osteoporosis. Earlier detection of changes in the structure and strength of bone and cartilage will facilitate earlier diagnosis, which is key to effective treatment and improved patient outcomes.

The function-dependent structure of bone trabeculae present challenges to measurement of mechanical properties. Extending preliminary work in applying BLS to measure the viscoelastic properties of bone and cartilage [89], Matsukawa *et al.* performed a comparative study with scanning acoustic microscopy (SAM) to determine the sensitivity of BLS in detecting anisotropy and decalcification, hallmarks of bone diseases [90]. Bovine trabecular bone samples were prepared in thin 30–150µm slices for transparency, with some treated with ethylenediamine- tetraacetic acid (EDTA) to model decalcification. Brillouin spectra were site-matched with SAM; a significant reduction in frequency shift (expressed as wave velocity) following decalcification was measured as well as higher signal intensity due to optical clearing effects of demineralization, demonstrating the applicability of BLS for characterizing bone quality in disease and healing processes.

Bone remodelling phenomena, particularly in the presence of implants, remain poorly understood due to the lack of means of evaluating implant stability and osseointegration in a non-invasive manner during healing post-implantation. To address this, Matthieu *et al.* investigated the capability of BLS in differentiating between mature and newly formed bone during post-implant healing based on viscoelastic properties and histological analysis [91]. Results indicated lower mineralization in newly formed bone compared to mature cortical bone, along with higher heterogeneity which reflects the initial lack of organization of ECM proteins during healing. This work confirmed the sensitivity of BLS in detecting mechanically-relevant compositional changes in bone during remodelling.

Akilbekova *et al.* similarly employed BM to investigate the load-dependency and viscoelastic properties of bone during healing. In [92], the efficacy of bone morphogenetic proteins (BMPs) and stem cell-loaded heparin-conjugated fibrin (HCF) hydrogels in the repair of critical bone defects in a rabbit model was assessed by measuring changes in BFS and linewidth, which are indicative of the rate of bone formation and healing processes at the site of the defects. Results were correlated to X-ray images, where Brillouin frequency shift and linewidth increased with formation of an organized endosteal callous, and showed that BMPs coupled with stem cells facilitated bone regeneration which is in line with previous findings.

Wu *et al.* also applied Brillouin microscopy to detect the loss of ECM proteins in cartilage as occurs in the pathogenesis of osteoarthritis. In [13], *ex vivo* porcine articular cartilage subjected to trypsin digestion for 4 hours yielded a 150 MHz decrease in BFS compared to untreated controls, confirming break down of structural ECM proteins; this was attributed to a small 4% increase in water content following enzymatic treatment.

Overall, these studies support the potential of BI as a non-invasive tool for characterizing mechanical properties of bone and cartilage during formation and healing, as well as a method for early detection of changes in these properties in disease.

### 4.3.3 Epithelium

Epithelium is the most abundant of tissues, which covers the body and forms the inner and outer lining of the majority of organs and internal cavities. Epithelial tissues are characteristically comprised of tightly packed cells on a basement membrane and perform a variety of functions including physical protection, secretion, absorption, excretion, filtration, diffusion, and sensory reception. Disruption of the mechanical integrity and physical properties of epithelia is implicated in a number of diseases. Palombo et al performed the first correlative Brillouin, Raman and FITR analysis of *ex vivo* Barrett's oesophageal tissue [56, 93] and demonstrated the capability of such a system for simultaneous site-matching of mechanical and chemical signature maps, which is valuable for detecting compositional and structural changes in oesophageal epithelia from chronic acid reflux as occurs in Barrett's oesophagus and other conditions such as cancer.

### 4.3.4 Cancers

Abnormal cellular functions and the uncontrolled division of these abnormal cells underpin a group of diseases referred to as cancer. Changes in gene expression, morphology and function in cancerous cells lead to distinct changes in the composition (biochemical, structural) and mechanical properties of tissues. An associated pathological changes in tissue stiffness would provide a diagnostic tool for early detection of cancerous and pre-cancerous cells and represent a potential application for Brillouin imaging. Troyanova-Wood *et al.* applied BM to study *ex vivo* samples of malignant melanoma in a porcine model and measured a distinct change in BFS between melanomas (8.55 ± 0.18 GHz) and healthy skin tissues (7.97 ± 0.02 GHz), with regressing tumors between this range (8.11 ± 0.07 GHz) [94], demonstrating the capability of BLS for differentiating between melanoma and normal skin, with potential for more rapid and non-invasive diagnosis of melanoma and other skin cancers.

While tissue stiffness is known to be important for cancer progression, metastasis and patient outcome [95, 96], there is increasing evidence that the mechanical interaction between cells and ECM may also influence cancer development and progression [97, 98]. However, these micromechanical properties and interactions are difficult to analyze with current technologies. Therefore, in addition to diagnostic applications, BI can be implemented to better understand the interplay between cell-matrix interactions in cancer development and progression for the development of more targeted and effective therapies.

### 4.3.5 Plaques

Disease processes can produce aggregates of matter known as plaques and are characteristic of respective pathological conditions; examples include dental plaque, amyloid proteins in Alzheimer's disease, psoriatic skin lesions and atherosclerotic plaque. As plaques accumulate during disease progression, they tend to become semi-hardened over time and disrupt the elasticity and functionality of the tissue *in situ*. Mattana *et al.* reported the first application of BM in combination with Raman spectroscopy to map the viscoelastic profiles of individual amyloid beta protein plaques in *ex vivo* histological sections of mouse hippocampus tissue [52] and showed, through spectral deconvolution algorithms, that amyloid plaques are comprised of a rigid core of β-folded (β-amyloid) proteins in a sheet conformation surrounded by a softer ring-shaped region richer in lipids and other protein conformations [99]. While a limitation with all optical methods is the penetration depth within a tissue, BLS can still be a valuable means of understanding the biomechanics of plaque formation in amyloidopathy in biopsy samples as well as evaluate the efficacy of treatments.

Atherosclerotic plaque is the hallmark presentation of atherosclerosis, where the endothelial wall of blood vessels is disrupted, allowing infiltration of lipoproteins that progressively accumulate and form a necrotic core [100]. Given that rupture of an atherosclerotic lesion leads to acute cardiac events and frequently death, early detection and treatment of rupture-prone plaques is key to patient survival. Meng *et al.* [101] and Antonacci *et al.* [14] investigated BM for giving quantitative measures of stiffness across rupture-prone atherosclerotic plaques, which are characterized by a thin, stiff fibrous cap covering a soft lipid-rich necrotic core within an arterial wall. They measured distinct differences in the mean frequency shifts of *ex vivo* plaque (15.79 ± 0.09 GHz) compared with normal vessel sections (16.24 ± 0.15 GHz) and showed a direct correlation between lower frequency shift for lipid-rich regions and higher for collagenous areas of the fibrous cap [14]. This study supports the combination of Brillouin spectroscopy with intravascular imaging to improve detection of vulnerable plaques and predict rupture potential in atherosclerotic patients.

### 4.3.6 Embryonic development

Mechanical forces have profound effects on cell behavior in adult tissues, but less is known about their roles during embryonic development where an embryo undergoes very dynamic changes in cell number, cell shape, cell contacts and migration to form functional organs and tissue structures [102]. Introduction of mechanical stimuli or chemical agents can disrupt highly sensitive processes within an embryo and therefore confound measurement, calling for a non-contact and label-free means of studying the biomechanical properties of cells and developing tissues during embryogenesis. Raghunathan *et al.* demonstrated the efficacy of Brillouin microscopy in combination with optical coherence tomography (OCT) to map and measure changes in stiffness of organs within a developing embryo [103], where OCT provides structural guidance for correlating Brillouin measurements. Using this technique, structures such as the neural fold and developing heart and closed neural tube could be successfully identified. Zhang *et al.* extended this work to investigate neural tube formation during embryonic development [60], further validating that Brillouin-OCT provides sufficient sensitivity and resolution to measure changes in stiffness of tissues and structures and its application in embryonic mechanobiology.

## 5. Data Acquisition, Processing & Analysis

### 5.1 Data acquisition and processing

Data acquisition, storage and processing in BLS experiments is typically done by custom-built scripts and software, developed by each research group to meet particular requirements [83] with the exception of GHOST software which is supplied with TFP interferometers as a part of the commercial product. GHOST software, produced by Table Stable Ltd in conjunction with the University of Perugia, is a versatile multi-channel analyzer equipped with curve fitting and calibration features.

Curve fitting is one of the most important signal processing routines necessary to obtain the Brillouin frequency shift $\Omega$ and the FWHM of the peaks $\Gamma$. Several models can be used to achieve this goal, with Lorentzian, Damped Harmonic Oscillator (DHO) and Gaussian models being most popular. The TFP interferometer and nonlinear coherent techniques are able of achieving high spectral resolution of approximately 1-50 MHz whereas typical spectral features from a sample are a few hundred MHz in width. Therefore, the latest generation of instruments can be used to resolve "true" line shapes unobstructed by the instrument function [3]. The hyperspectral analysis to decompose overlaid spectral features of Brillouin peaks has recently been applied to study amyloid-beta plaques in Alzheimer's disease brains [99]. This technique, previously found to be useful to extract information in Raman imaging, has strong potential in Brillouin imaging of heterogeneous environments within the scattering volume such as majority of biological tissues and cells.

### 5.2 Models to interpret Brillouin scattering data in biological matter

Historically, the physics of BLS was first applied to understating photon-phonon interactions and structural relaxations in gases, liquids and solid state crystalline materials, in which the stress-strain relationship can be described by the models of an ideal gas, a Newtonian fluid or a linearly elastic solid, respectively [104]. Recent application of BLS to study biomechanics of cells, tissues, matrix-like biomaterials and whole organisms, has challenged physicists and life scientists, calling for new theories that can explain BLS data and connect findings with other available methods of micromechanical testing such as AFM micro-indentation, optical tweezers, and micro-rheology. The search for new hybrid models that can explain BLS measurements in biological matter is one of the greatest challenges in the field, and the solution to this problem is much needed in order for the technology to be recognized by the rest of scientific community and translated to commercial applications. There have been a number of proposals that attempted to interpret BLS data by adopting viscoelastic or poroelastic models and we briefly review these in the next sections.

### 5.2.1 Viscoelastic model

Viscoelasticity is the property of materials that exhibit both elastic and viscous characteristics when undergoing deformation. This dual nature can be described by setting elastic moduli to be complex and frequency-dependent. The complex longitudinal modulus $M^*(\omega) = M'(\omega) + M''(\omega)$ is then composed of the storage $M'(\omega)$ and the loss $M''(\omega)$ components where ω is the driving frequency of mechanical deformation. The former describes elastic material response to deformation, and the latter refers to energy dissipation due to viscous effects. According to this model, Brillouin peaks can be reproduced by a damped harmonic oscillator (DHO) function

$$I(\omega) = \frac{I_0}{\pi} \frac{\Gamma\Omega^2}{(\omega^2-\Omega^2)^2+(\Gamma\omega)^2} \tag{5}$$

convolved with the measurement instrument function. The Brillouin frequency shift Ω and line width Γ derived from this model yield the storage and loss moduli at the peak frequency $\omega = \Omega$

$$M'(\Omega) = \frac{\rho}{q^2}\Omega^2 = \rho V^2, \tag{6a}$$

$$M''(\Omega) = \frac{\rho}{q^2}\Omega\Gamma = \eta\Omega, \tag{6b}$$

where $\eta = \frac{\rho}{q^2}\Gamma$ is the longitudinal kinematic viscosity.

The single peak analysis shown above is the standard data analysis method in BLS applied to biology and biomedical fields. Nevertheless, there are a few other functions adopted for fitting Brillouin data that we have already discussed in Section 3.3.

The important consideration is the frequency dependence of the storage and loss moduli in Eq. (5) and (6). BLS experiments probe acoustic phonons in GHz frequency range and hence the measured frequency shift Ω and line width Γ correspond to the material response at supersonic frequencies. Frequency dispersion of viscoelasticity manifests itself in drastically different response to stress perturbations below and above the structural relaxation frequency $\omega_r$ which is unique for each material. An intuitive explanation for material response below and above the relaxation frequency $\omega_r$ was presented by Palombo and Fioretto in their recent review [3].

It is worth noting that the supersonic frequency regime relevant to BLS experiments typically corresponds to material response above $\omega_r$. Under GHz perturbation, the material does not have enough time to relax and appears stiffer (higher values of $M'$) compared to the perturbation of the same amplitude but much lower frequency (Hz-kHz), typical of micro-rheology or micro-indentation measurements. Because the two groups of methods, low and high frequency, assess system's relaxation dynamics in two limits, these methods should be viewed as complementary to each other. It has been shown, however, that changes in storage and loss moduli resulting from pathological processes or drug treatment do correlate across the broad frequency range, thus showing similar trends at Hz and GHz frequencies [2, 11]. This, however, is merely a rule-of-thumb and correlations should be verified for each specific material platform and experimental conditions.

*5.2.2 Poro-elastic model*

Recently a new interpretation of BLS results in tumor spheroids has been proposed in which the poro-elastic nature of some biomaterials (e.g. hydrogels) and tissues is considered [11]. A poro-elastic material can be represented by an elastic porous frame immersed in a viscous fluid. The stress tensor $\sigma_{ij}$ and mean fluid pressure $p$ are estimated on the basis of a minimal Darcy-scale poro-elastic model as a function of the strain tensor $\varepsilon_{ij}$ and fluid content $\xi$ written in index notation as

$$\sigma_{ij} = K_u \varepsilon_{kk} \delta_{ij} + 2\mu \varepsilon_{ij} - \frac{K_u - K_d}{\beta} \xi \delta_{ij}, \tag{7}$$

$$p = \frac{K_u - K_d}{\beta^2} (\xi - \beta \varepsilon_{kk}), \tag{8}$$

where $K_u$ and $K_d$ are the undrained and drained bulk moduli, respectively, $\mu$ is the shear modulus of the drained material and $\delta_{ij}$ is the Kronecker delta.

The fluid is typically allowed to drain through open pores of the solid frame under application of stress. Depending on the stress rate and critical time required for fluid to drain out, different scenarios can occur. The long term stress-strain response of tissues simulates the drained case where the fluid is allowed to flow freely ($p = 0$) and gives access to $K_d$. At the high-frequency limit, however, the fluid has no time to drain and becomes trapped in the pores ($\xi = 0$). A critical frequency $\omega_c = 2\pi f_c$ differentiates between these two limiting regimes. It has been found that for hydrogels this frequency is expected to be in the range $f_c = 1 - 200$ GHz [11]. This suggests that photon-phonon interaction underlying BLS can probe the poro-elastic system dynamics close to the threshold between drained and undrained regimes.

It has been suggested that when approaching $f_c$, variations in the longitudinal modulus and hence the Brillouin frequency shift can be interpreted as a variation in a volume fraction of circulating water $\phi$, so that $M' \sim M'_w / \phi^3$, with $M'_w$ being the longitudinal modulus of fluid phase [11]. According to such interpretation, the increase in $M'$ towards the center of the tumor spheroid could be related to the reduction in the amount of circulating water [11].

In general, the link between local hydration and the Brillouin frequency shift is well known in BLS community and has been studied earlier in relation to isolated collagen and elastin fibers [51], hydrogels [30] and cornea [44]. Both longitudinal $M$ and Young's $E$ moduli are affected by fluctuations in tissue hydration and at GHz frequencies the effect of hydration becomes prominent for $M$ in particular since fluid has no time to drain and behaves as solid.

*5.2.3 Two-phase mixture model for highly hydrated biomaterials*

In 2018 a seminal work was published addressing the question of local hydration and its effect on the longitudinal and Young's moduli in hydrogels with liquid content above 90% [30]. The authors conducted a study to answer one of the important questions in BI: is correlation between $M$ and $E$ universal and valid for all biomaterials? If this were found true, $\Omega$ could be interpreted as a measure of stiffness, the term typically associated with $E$. This timely question required an answer since the only evidence confirming the relation of $\Omega$ to stiffness was based on a phenomenological formula

$$\log M' = a \log E' + b \tag{9}$$

obtained in a set of experiments using porcine and bovine lens tissues [43].

Stiffness is typically understood as the amount of resistance offered by an elastic body to deformation. This definition of stiffness is very general and does not include details of deformation nor the boundary conditions applied. In Section 2.1 we have already discussed the differences between $M$ and $E$ being in the boundary conditions. Namely, in the definition of $M$ the deformation is constrained to a single axis, same as the direction of applied stress, and hence the meaning of $M$ is more closely linked to the material compressibility. $E$, in contrast, is an unconstrained modulus with matter being free to move in all directions upon the application of stress. This seemingly small difference in the boundary conditions creates dramatic consequences for biomaterials and tissues which are soft and highly hydrated. Since it requires much larger force to compress a tissue full of fluid than to deform it, the difference between $M$ and $E$ can be many orders of magnitude. There is also no reason to believe that $M$ and $E$ are universally connected by a simple proportionality rule as Eq. (9) due to wide variety of possible deformation scenarios and level of structural and composition complexity inherent to tissues and cells.

Wu *et al.* set to examine the universal validity of Eq. (9) in a set of experiments with hydrogels of two compositions, i.e. polyethylene oxide (PEO) and polyacrylamide (PA). Since swelling is a common feature of hydrogels (as well as tissues and cells), the gel hydration level $\epsilon$ was measured and kept as an independent parameter. The hydration level in both types of gels was varied between 90% and 100%. Both $M$ and $E$ were influenced by $\epsilon$, but more importantly no single law correlation was found between $M$ and $E$. For example, the same value of $M$ was measured for PEO hydrogels with molecular weights of 1, 4 and 8 MDa but an independent measurement of $E$ using an unconfined compression test indicated a five-fold difference in the Young's modulus between gels of 1 and 8 MDa.

To explain the results of Wu *et al.*, one can adopt the poro-elastic model with fluid content of $\epsilon > 0.9$. Due to high fluid content, Eq. (7-8) can be significantly simplified, leading to a biphasic model of hydrogels proposed in earlier works by Hosea [105] and Johnston [106]. According to that the aggregate compressibility of the gel, $\beta_{agg} = 1/M_{agg}$ is a superposition of individual compressibilities of its two phases, a fluid phase with the bulk modulus $M_f$ and a solid one with $M_s$

$$\frac{1}{M_{agg}} = \frac{\epsilon}{M_f} + \frac{1-\epsilon}{M_s}. \tag{10}$$

The linear relation described by Eq. (10) is clearly an approximation and does not take into account many important factors of hydrogel viscoelasticity, e.g. any interaction between fluid and solid fractions, contribution of bound fluid to the polymer networks, the internal geometric structure of the pores, or the network topography. The biphasic model, however, proves to be quite useful in the explanation of Brillouin scattering data from highly-hydrated materials such as hydrogels [25] and cornea [30].

## 6. Challenges for clinical application of Brillouin Imaging

Brillouin imaging is rapidly becoming useful as a research tool for studying structure-function relationships in biological cells and tissues. Although in a nascent stage, the field is already characterized by a great diversity of approaches and custom equipment. While increased accessibility to optical components has given rise to rapid development of new techniques and system configurations, such multiplicity raises a particular risk. Namely, the scientific relevance of any Brillouin light scattering-based study is as good (or lacking) as the alignment between (1) the research question, (2) system capabilities, (3) the measurement protocol, (4) the assumptions made about the studied sample, and (5) the models and algorithms used to analyze the raw Brillouin spectral data. Furthermore, Brillouin imaging in itself is not without its challenges, presented categorically in the following sections.

## 6.1 Sample preparation

Biological tissues have complex microarchitectures and are not optically clear. Clearing reagents and protocols have been proposed, however, clearing treatments and even conventional tissue fixation methods can be expected to disrupt and alter the physical properties of cells and tissue components. The effects of the sample preparation procedures need to be investigated in the context of BLS and caution should be taken in the interpretation and extension of the results measured from treated *ex vivo* samples to *in vivo* and *in situ* properties. In addition, given that some BI studies are performed on fixed or *ex vivo* samples, it is critical that the possible impact of sample preparation, handling and storage techniques on subsequent mechanical properties be carefully established. Live *in vivo* or *in vitro* imaging, or other techniques that negate the need for chemical or physical modification of a sample will allow measurement of more physiologically-accurate information.

## 6.2 Analysis and interpretation of Brillouin data

Elucidation of Brillouin data sampled across a focal volume can become non-trivial in heterogeneous tissue samples. Nevertheless, it is possible to correlate Brillouin frequency shifts with different components within a heterogeneous sample using multispectral reconstruction numerical techniques and additional imaging modalities, e.g. Brillouin-Raman imaging. Furthermore, it is likely that signal deconvolution algorithms will be continually refined and optimized to enable conclusive, site-matched measurements.

While Brillouin frequency shift has increasingly been used interchangeably with 'stiffness' by biologists, other material parameters are required in order to calculate longitudinal modulus, an accepted mechanical metric, from Brillouin frequency shift. These include refractive index and density, which are challenging to measure for composite and heterogeneous materials such as biological matter or even an individual cell. While several methods for measuring refractive index exist, e.g. digital refractometers, there must be a means of conveniently measuring refractive index in a site-matched manner across 3D samples to enable accurate approximation of longitudinal moduli with sufficient spatial resolution. One way of overcoming this is to directly measure refractive index within a Brillouin system. Fiore et al recently devised a means to simultaneously measure refractive index using a dual co-localized BLS configuration [107].

As the Brillouin community grows, there could be a collaborative effort to develop an open-source library of accepted values for different tissue types and components to simplify modelling and extraction of longitudinal moduli from Brillouin spectra. Otherwise, Brillouin frequency shift and linewidth in themselves could potentially become accepted standalone unitized metrics with implicit/direct correlation to longitudinal modulus wherever relative (not absolute) measures of stiffness suffice e.g. for assessing changes in cell or tissue stiffness post-treatment.

## 6.3 Standardized protocols for BLS measurements

At present, the design of Brillouin systems is in a developmental stage with the aim to optimize and refine particular aspects of BI as a technique, e.g. resolution, or exploratory in developing further techniques e.g. combinative systems. However, as hardware and system configuration determines capability and results, there exists a need for a comparative study to identify system variability in order to enable meaningful comparison of results obtained from a diverse range of system designs; this could be done by say, measuring a set of standard samples (distilled water, ethanol, selected cell lines) across different instruments as a first step towards standardization.

Once instrument variability is addressed, the consensus on standardized protocols for measuring different categories of biological samples of interest will facilitate the community to achieve reproducible measurements and allow for more ready comparison of data between researchers.

These protocols can be further developed and refined for BI measurements of tissues in order to first determine reliable statistical thresholds of 'normal' and 'abnormal' Brillouin results for each tissue type, using parameters that can be clearly differentiated at the 95% confidence level (or better), before extension to disease detection and clinical diagnosis. Finally, a clear guide for selecting Brillouin components, e.g. laser wavelength, laser power and spectrometer type to satisfy majority of biological and biomedical experiments could be useful for non-experts in the field.

## 7. Conclusion

Owing to its inherent advantage of non-contact measurement, Brillouin imaging offers real capabilities in providing unique insights into the physical properties of biological cells and tissues. However, use of Brillouin imaging in a diagnostic capacity is still in its infancy with several issues requiring attention. Firstly, the instrumentation itself will need to become more standardized, economical and smaller as well as redesigned with clinicians as the end-user, i.e. more ergonomic with user-friendly data acquisition and handling interfaces. The issue of mechanical robustness of any machine that includes a Fabry-Perot etalon needs to be considered due to the latter containing relatively delicate moving components. Protocols for sample preparation and measurement will need to be standardized and consolidated across a range of instrument designs. Reliable statistical envelopes for the BFS maps of 'normal' or 'abnormal' tissues will need to be determined for selected medically important pathologies through larger-scale clinical trials in order to develop diagnostic algorithms which can then be packaged as software to complement commercialized clinical systems.

Although we have shown in this review that BLS-based imaging and spectroscopy can offer useful data for a great range of medical studies, it will only be feasible to commercialize the techniques for a few conditions at first, and then likely only in the form of *ex vivo* biopsies for non-surface/deep tissues. Tumor detection and growth, characterization of systemic sclerosis, or monitoring of plaque formation, as examples, may offer suitable targets for the initial transfer of this technology to clinical settings. The ultimate challenge, however, is to combine this technique with endoscopy for routine *in vivo* measurements in clinical settings. Given the valuable insights that Brillouin imaging can produce and how rapidly the field has progressed in recent years, we have little doubt that this goal will be reached in time.

## Statement of Contribution

CP devised the direction and theme of the review, designed the main structure and outline of the document, wrote sections 3.3, 4 & 6, helped write sections 1, 3.2 and 7 and provided general editing. JC edited the document, provided general editing and feedback on section 4. MC edited the document, provided critical insights and some additional references, and helped write sections 6 & 7. IK wrote sections 1-3 & 5, helped write section 6 and 7, provided general editing and contributed with critical planning and discussion of the manuscript.

## Acknowledgements

This work was supported by the Australian Research Council through Discovery project DP190101973.